\documentclass[pre,aps]{revtex4}
\usepackage{graphicx}
\usepackage{array}
\usepackage{url}

\usepackage{verbatim}

\newcommand{\req}[1]{(\ref{#1})}
\newcommand{\be}{\begin{equation}}
\newcommand{\ee}{\end{equation}}
\newcommand{\bea}{\begin{eqnarray}}
\newcommand{\eea}{\end{eqnarray}}

\newcommand{\pr}[1]{\left(#1\right)}

\newcommand{\avg}[1]{\langle{#1}\rangle}

\newcommand{\BE}{\begin{eqnarray}}
\newcommand{\EE}{\end{eqnarray}}
\newcommand{\BEn}{\begin{eqnarray*}}
\newcommand{\EEn}{\end{eqnarray*}}
\newcommand{\barr}{\begin{array}}
\newcommand{\earr}{\end{array}}

\newcommand{\bit}{\begin{itemize}}
\newcommand{\eit}{\end{itemize}}
\newcommand{\bc}{\begin{center}}
\newcommand{\ec}{\end{center}}
\newcommand{\ben}{\begin{enumerate}}
\newcommand{\een}{\end{enumerate}}

\begin{document}
%
\title{Fat tails, long memory, maturity and ageing in open-source software projects}

\author{Damien Challet}
\affiliation{Institute for Scientific Interchange, via S. Severo 65, 10113 Turin, Italy}
\affiliation{D\'epartement de Physique, Universit\'e  de Fribourg, P\'erolles, 1700 Fribourg, Switzerland}
\email{challet@isi.it}
\author{Sergi Valverde}
\affiliation{ICREA-Complex Systems Lab, Pompeu Fabra University, Dr. Aiguader 80, 08003 Barcelona, Spain}
\affiliation{Centre de Recherches sur la Cognition Animale, CNRS-UMR 5169, Universit\'e Paul Sabatier,
118, route de Narbonne, 31062 Toulouse Cedex 04 France
}
\email{svalverde@imim.es}


%



\begin{abstract}
We report activity data analysis on several open source software projects, focusing on time
between modifications and on the number of files modified at once. Both have fat-tailed distributions,
long-term memory, and display systematic non-trivial cross-correlations, suggesting that quiet periods
are followed by cascading modifications. In addition the maturity of a software project can be measured from the exponent of the  distribution of inter-modification time. Finally, the dynamics of a single file displays ageing, the average rate of modifications decaying as a function of time following a power-law.
\end{abstract}


\maketitle

%

\section{Introduction}



The time between consecutive events
observed in many human activities is neither Poissonian nor Markovian, but exhibits bursts of rapidly occurring events separated by long periods of inactivity. The distribution of interevent times follows heavy-tailed distributions
\cite{JohansenActivity,BarabasiActivity,VazquezActivity,BarabasiActivity2,CommentOnBarabasi}.

Although activity patterns resulting from an aggregated behaviour, such as financial markets, have
been studied for a long time, recent work focused on the individual behaviour. It has been shown
that in some cases non-Poissonian behaviour is not only a by-product of human interaction, but can be
traced back to individuals.  Some remarkable examples are web surfing and
e-mail communication \cite{JohansenActivity,BarabasiActivity,Fifteen} .  An important issue is the relationship between individual behaviour and aggregate activity, which we shall investigate with the help of an important, yet under-explored,
archive of human activity:  open-source software ({\sc oss}) development.  Distributed communities of programmers
develop open-source software projects \cite{ValverdeEmail06}, where several programmers change simultaneously possibly more than one piece of software.

 In the past, indices of global development activity have been devised to assess
the stability of a software system. A popular measure, called software volatility, measures the
number of enhancements per unit of time over a specified
time frame \cite{BankerSlaughter2000}.  High volatility is often associated to high
 maintenance costs; according to this point of view, when the volatility exceeds some threshold it may be more economical to rewrite the entire system from scratch instead of maintaining an aged (and unstable) software system \cite{Chan1996, Heales02}. We show here however that the probability distribution of scaled waiting times of the software projects converges towards a universal function, providing {\em a contrario} a measure of software maturity.

We also investigate the relationship between individual programmer behaviour and global project
activity by means of a detailed statistical analysis of both the timing and size of individual
actions.
  Unlike previous studies (e.g. \cite{Heales02}),
we do not attempt to make any distinction between different types of modifications.


The vast majority of software projects keep track of every single change and its author,
using various so-called version control systems. {\sc CVS} (concurrent versioning system)
is a frequently used system in open source projects.
In order to avoid costly information losses, the {\sc CVS} keeps
each programmer apart by managing multiple revisions for
each project file. These features makes the {\sc CVS} an invaluable source of
information about software evolution and programmer activity patterns.
We have analyzed the CVS databases of six large-scale open-source projects from their creation date until November 2005:
{\sc Mozilla}, {\sc Apache}, {\sc FreeBSD}, {\sc OpenBSD}, {\sc NetBSD},
and {\sc PostgreSQL}.

\section{Analying CVS logs}

\begin{figure}[htbp]
\begin{center}
\includegraphics[width=.45\textwidth]{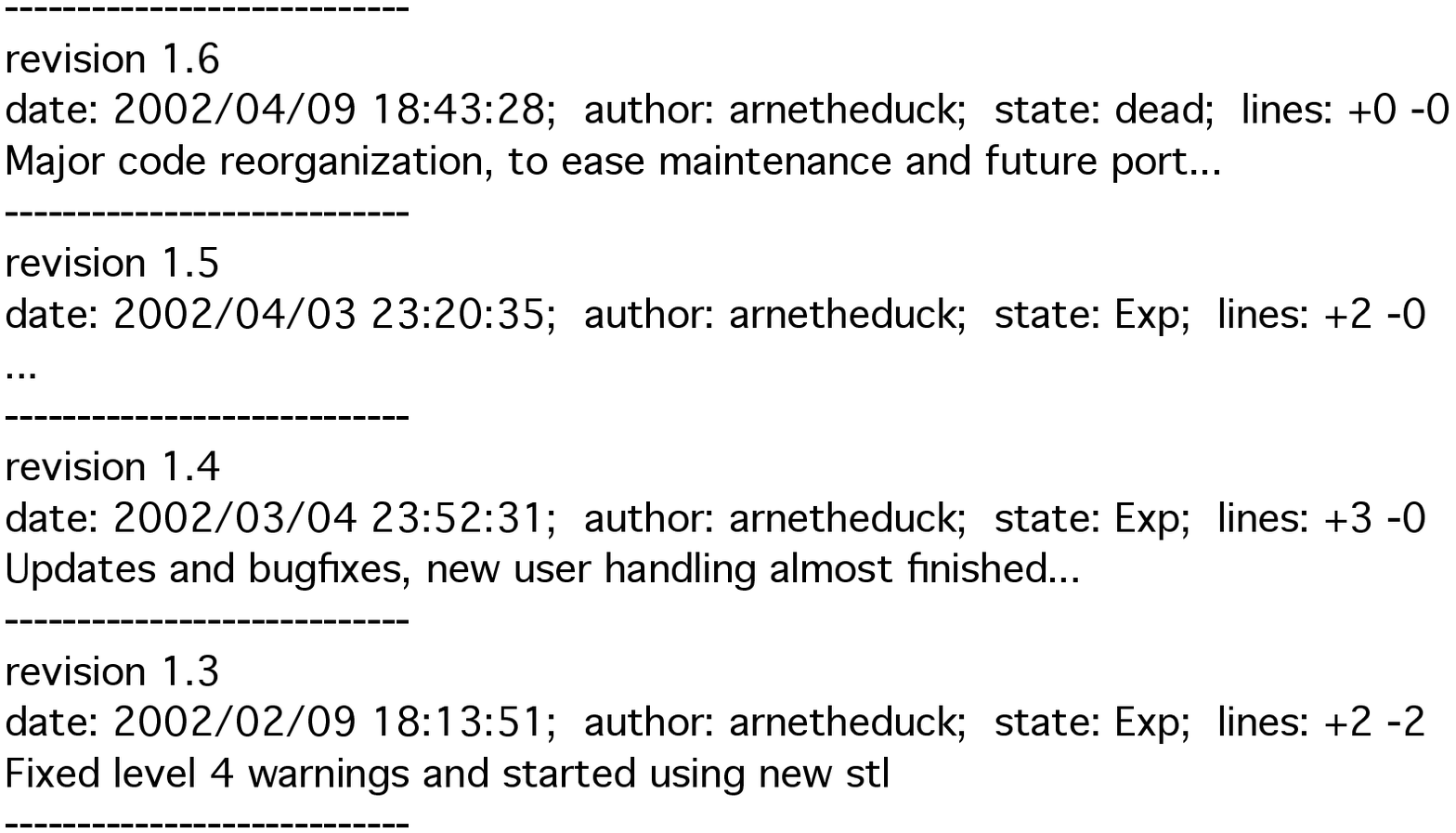}
\caption{ An example of CVS log entry generated by the {\em cvs log}  command for a file in the {\sc dcpp} project. Each entry describes a single file revision, indicating revision number,
date,  author's name, state, number of lines added and removed and a brief description.
The raw CVS log is human-readable but difficult to analyze by automated means.}
\label{cvs}
\end{center}
\end{figure}

Analyzing operational logs of OSS project developments stored in CVS code repositories requires sometimes delicate ad-hoc pre-processing, without which spurious effects can easily spoil the results in an uncontrollable way. Indeed, while CVS stores all of its information in a text file describing in a human-legible way the nature of each modification (see Fig. \ref{cvs}), there is no standard format definition; it takes therefore some efforts to extract valuable and trustable information from CVS log files.

\begin{figure}[htbp]
\begin{center}
\includegraphics[width=.45\textwidth]{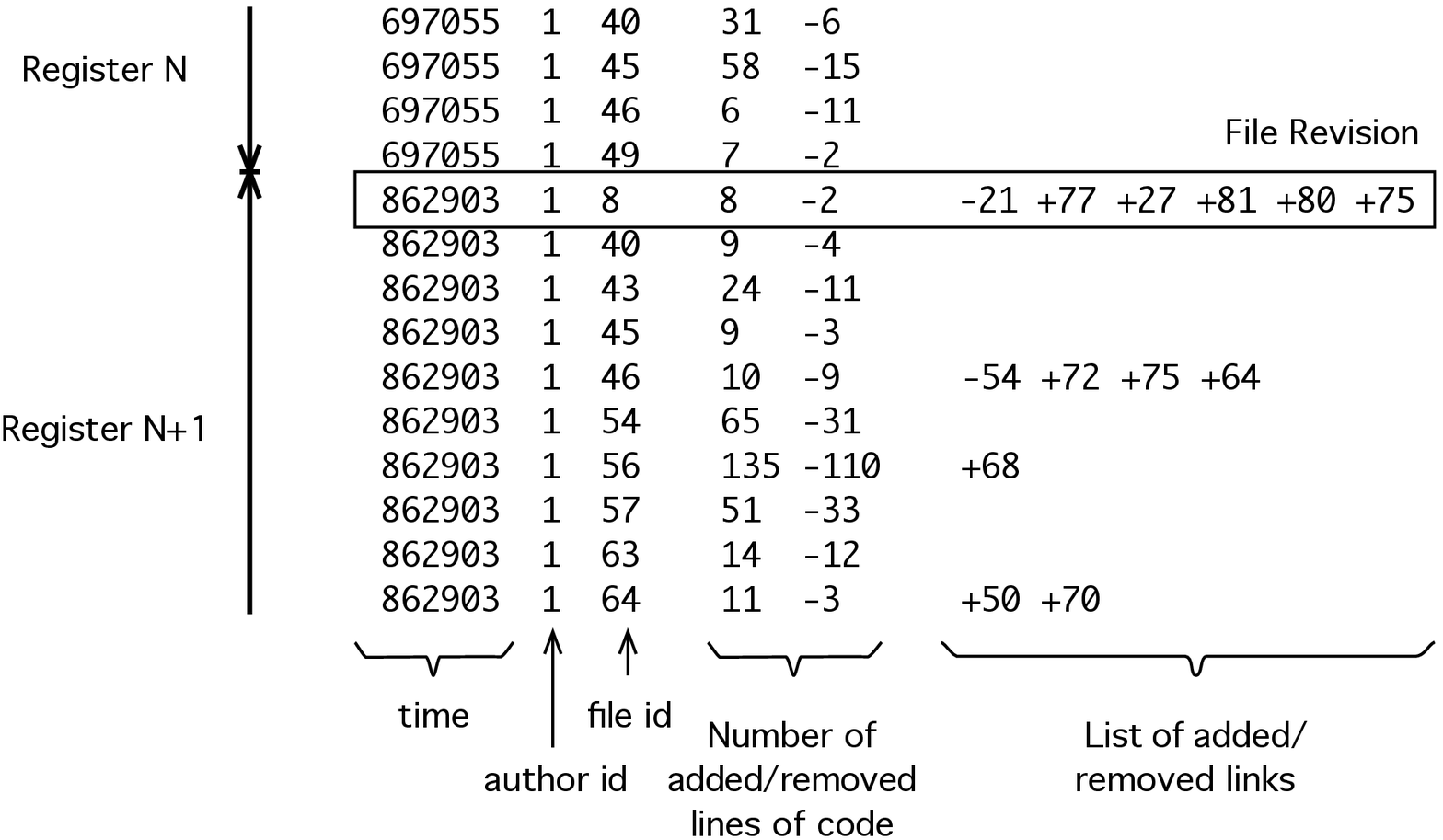}
\caption{ An example of {\em event log} file generated from
the CVS log file of the {\sc DCPP} project. This file format allows for simpler numerical
analysis of development patterns.  }
\label{event}
\end{center}
\end{figure}

We parsed CVS logs and generated
synthetic files (thereafter {\em event logs}) which are more amenable to analysis (see Fig.\ \ref{event} for an example).   Every entry in this file describes a single file revision and
provides information about the event time, author's identifier, unique file identifier,
number of changed lines of code (added and/or removed) and a detailed list of link changes (added and/or removed) (see below).

\subsection{Filename changes}

\begin{figure}[htbp]
\begin{center}
\includegraphics[width=.5\textwidth]{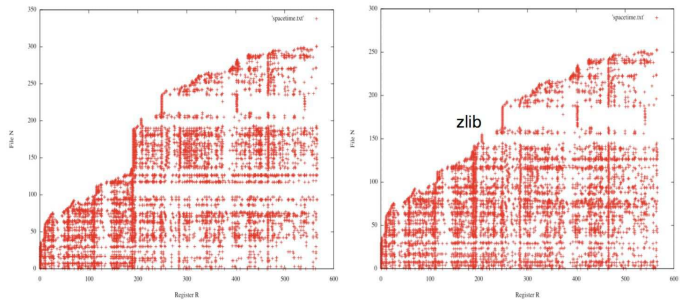}
\caption{Activity map of {\sc dcpp} before and after filename aliasing; each point represents a single file change. In the right hand side figure, the label {\sc zlib} points at a stream of modifications carried out on an auxiliary software library.}
\label{rename}
\end{center}
\end{figure}

Unfortunately, CVS logs have a number of shortcomings. For instance, CVS does not handle file rename or file moves, making it difficult to rebuild faithfully the evolution of software structure. Instead, every renamed file generates a new file in the CVS, resulting in two entries for the same file and giving rise to specific patterns in the space-time map (see Fig.\ \ref{rename}).

There is no general easy way to implement name aliasing. Two simple approaches can address this problem: (i) to deal only with the set of CVS registers  before the actual file rename
takes place or (ii) to fix the CVS manually by providing a list of filename aliases.  Which scheme to choses depends on the size of CVS log. For instance, (i) is a suitable approach for very large CVS logs with few file renames. On the other hand, the small size of some CVS logs allows the file moves to be fixed by human inspection. Moreover, one cannot discard registers from small CVS logs without affecting the statistics (for instance, the long tail of the distribution of size of changes).

Some peculiar patterns in the space-time diagram appear as groups of files  change together in a remarkably synchronized fashion (see the upper region of the plot in Fig. \ref{rename}b). These patterns represent synchronization events between auxiliary software components (i.e.,  libraries or frameworks) and the core software application . The external software library evolves in parallel and  is maintained by external software team. However, the interaction between these systems is often asymmetric. Changes to the auxiliary software library are exogenous perturbations in the core application; interaction rarely happens the other way around. A statistical way of detecting such processes was proposed by one of us \cite{ValverdeCrossover}.

\subsection{Software structure}

Because of software reuse, files link to each other, thereby defining a network of dependency which plays an important role in software dynamics (see e.g. \cite{ValverdeOptimalDesign,ValverdeLog05,Myers,CL04}). Visualizing structural changes helps understanding the microscopic dynamics.
However, CVS log files do not contain any structural information.
\begin{figure}[htbp]
\begin{center}
\includegraphics[width=.45\textwidth,height=0.43\textwidth]{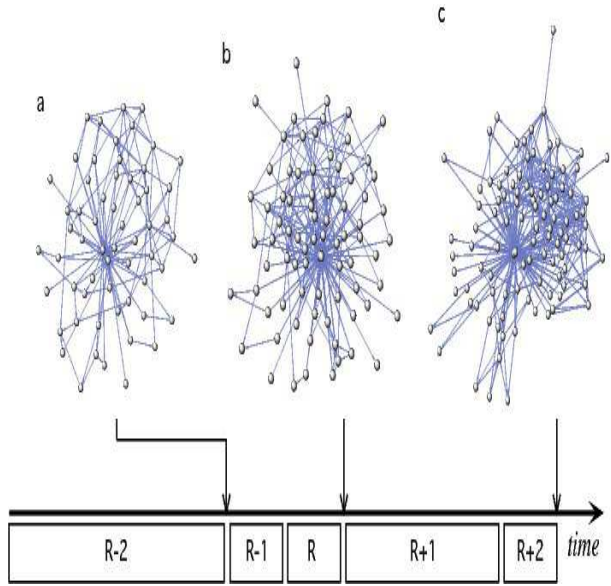}
\caption{ Mapping between software networks and the stream of CVS registers. Here, software network
$a$ maps to register $R-2$, software network $b$ maps to registers $R-1$ and $R$ and finally,
software network $c$ maps to registers $R+1$ and $R+2$. }
\label{decorate}
\end{center}
\end{figure}

The first step is to obtain enough source code versions for a given project. Since this can generate a large volume of information, one has to limit oneself to a subset of all versions for large OSS projects.  Then, it is easy to define a mapping between the sequence of software networks and the time evolution of CVS registers (see Fig.\ \ref{decorate}) and to add information about insertion/deletion of new files and links.




\section{Data analysis}

From each {\sc CVS} database history, we analyze the time, the author
and the number of modified files of each source code alteration.  Each development
history comprises a  number $M$ of modification registers.  Hence, the $i$-th
modification was at time $t_i$ by programmer $a_i$ and concerned $s_i>=1$ files,
$1 \le i \le M$.  Modification attributes can be obtained as the sum of individual
contributions. Lets $c_{jk}(t) = 1$ if programmer $j$ modifies file $k$ at time $t$ and
$c_{jk}(t) = 0$ otherwise. Then,

\begin{equation}
S_i  = \sum\limits_{1 \le k \le N_f } {c_{a_i k} (t_i )}
\label{e2}
\end{equation}

measures the number of changed files at time $t_i$, thereafter called modification size.

\subsection{Interevent and modification size distributions}

We will be interested in the interevent
distribution $P(T)$ of between two consecutive modifications
  The time elapsed between two consecutive
modifications $i$ and $i-1$  is denoted by $T_i=t_i-t_{i-1}$.  It reflects a maturation process at
a particular stage of software development and depends on many factors, such as
the cognitive capabilities of programmers \cite{Psychological},  team composition
\cite{Clark1998}, software usage
  or deadlines.
We report the distribution of interevent times at three different scales: project,
individual developers, and files. In order to differentiate them, we adopt the following notation: $T_i^{(f)}$ is the time interval
between two consecutive modifications of file $f$,  $T_i^{[a]}$ , the  time interval  between two consecutive modifications  made by author $a$, while the absence of a superscript denotes the global project level. Therefore, we focus on $P(T)$, $P(T^{(f)})$, $P(T^{[a]})$,  $P(S)$ and $P(S^a)$, and how they are related to each other.

Previous work produced plots {\em en masse} of the number of files modified by a given author,
or the number of modifications contributed per author, without further analysis such as
fitting, scaling or discussion of stationarity \cite{CVS-Spain}. A related work consisted in
measuring the distributions of number of lines of code added or deleted per modification,
which are both clean power-laws with exponent around $-3/2$ \cite{PismakSoftware}, whose origin is still unclear. Previous work has compared {\sc Apache} and {\sc Mozilla} in a qualitative way \cite{ApacheMozilla}.
 Finally, recent work studied the size of modifications and carried out a detrended fluctuation analys \cite{WuHolt}


\subsection{Time series analysis: caveats}

Human beings have intrinsic time scales. Despite this obvious fact, one finds frequent bursts of  modifications from the same programmer occurring with super-natural high frequencies. This comes from the way some programmers submit their modifications to CVS with automated scripts, for instance every
5 seconds. One must therefore coarsen the time series so as to remove  non-human dynamics by merging the  modifications $i$ and $i+1$ separated
 by less than $\delta T$ seconds, i.e., $t_{i + 1}  - t_i  < \delta T$; the influence of $\delta T$ is discussed thereafter.

The other fundamental problem is that the dynamics of software projects is seldom stationary during their whole histories. Indeed, one characteristics of the successful large-scale open-source projects analyzed here is to gradually attract the attention of the public and of companies, resulting into an increasing number of programmers, denoted with. $N_a(t)$. In addition, in any software project,  the number of files $N_f(t)$ increases as a well. Since we are first focusing on global activity patterns, this may not be a problem if for instance the activity of each programmer, or the rate of modifications of a given file decreases in a related manner.

\begin{figure}
\centerline{\includegraphics*[width=0.45\textwidth]{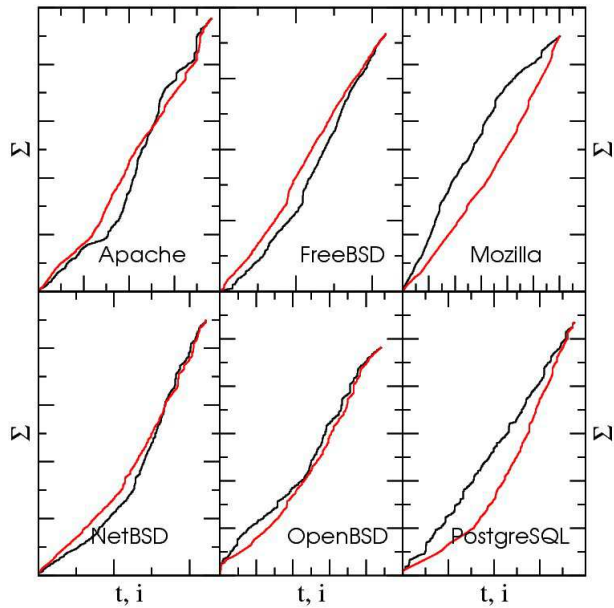}}
\caption{Cumulative number of modified files $\Sigma$ as a function of time $t$ (black lines)
and of modification number $i$ (red lines). The units of $t$ and $i$,  $\Sigma(t)$ and $\Sigma(i)$ have been chosen so as to make the end of the time series coincide.}
\label{fig:nmods_vs_t_i}
\end{figure}

 A simple and effective way to measure the change of programming pattern is to plot measures of
cumulated activity. The cumulative number of modified files as a function of the time $t$,
defined as

\begin{equation}
\Sigma(t)=\sum_{t_i\le t}S_i
\end{equation}

is a good candidate, as its slope is equal to the rate of modified files per unit of time and
reflects therefore the global activity of the project. Fig \ref{fig:nmods_vs_t_i} reports
$\Sigma(t)$ for six software projects. Some of them have a relatively constant rate of
modification after a transient period ({\sc FreeBSD}, {\sc NetBSD}), while {\sc PostgreSQL}
has a remarkably constant modification rate. {\sc Apache} has a more erratic behaviour \footnote{This is in line with the qualitative study of Ref. \cite{ApacheMozilla}},
while {\sc OpenBSD} experiences various episodes of high and low activity, the activity reducing
lately. Finally, {\sc Mozilla}'s activity has been decreasing for much of its history, which can
be a sign of software maturing.

$\Sigma(t)$ does not characterize entirely the activity of a project.  Let us define a related quantity, the cumulative number $\Sigma(i)$ of modified files as a function of modification number $i$:

\begin{equation}
\Sigma(i) = \sum_{ j\le i}S_j
\end{equation}

The slope of $\Sigma(i)$ is equal to the average modification size $\avg{S}$ and therefore provides an additional criterion of stationarity.  The example of {\sc PostgreSQL} is striking in this respect:
although the slope of  $\Sigma(t)$ is remarkably constant, $\Sigma(i)$ increases superlinearly for at
least half of the time series before reaching a constant slope, meaning that the number of modifications
per batches has increased. This increase, although usually less spectacular, is seen in almost all the projects. This could reflect the evolution of the software structure: changes are likely to propagate to nearest neighbors, hence $\avg{\Sigma(i)}$ follows in part the evolution of the average number of nearest neighbours; and indeed one of us \cite{ValverdeCrossover}, found an initial increase of propagation of changes to nearest neighbors and then a saturation, confirming the above interpretation. The only exception here is
{\sc Mozilla} whose modification rate decreases at the end of its time series,
while the average number of modified files shows no sign of decrease; this is
the signature of a genuine slowdown of the development.

In short, a transient state is generally present at the beginning of projects histories
and a steady state is reached after some time. {\sc Mozilla} is clearly not in a steady state
in the last part of the history considered here, its modification rate having slowed down
significantly. {\sc OpenBSD} behaviour changes much with time, and {\sc Apache} is erratic.
At any rate, the history of a software project is generally not very smooth, the first part of
its development is different from the following ones, hinting at how crucial it is to split the
timeseries into several parts, which will be confirmed in the next subsections.


\begin{figure}
\centerline{\includegraphics*[width=0.45\textwidth]{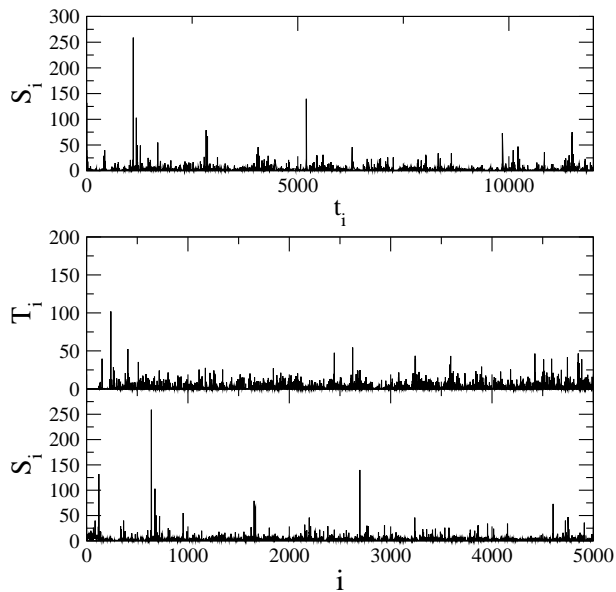}}
\caption{Activity patterns: number of modifications vs time of modification (upper graph),
time interval between two modifications (middle graph) and number of modified files (lower graph)
as a function of the modification number ({\sc Inkscape}).}
\label{fig:T(t)}
\end{figure}


Plotting $S_i$ as a function of $t_i$,  and $T_i$ and $S_i$ as a function of $i$ reveals a highly
non-trivial temporal structure (Fig \ref{fig:T(t)}). In particular, this figure displays two distinctive features found in all the datasets we studied: non-Gaussianity and clustered activity, which we will characterize in details.



\subsection{Time between modifications}

\begin{figure}
\centerline{\includegraphics*[width=0.45\textwidth]{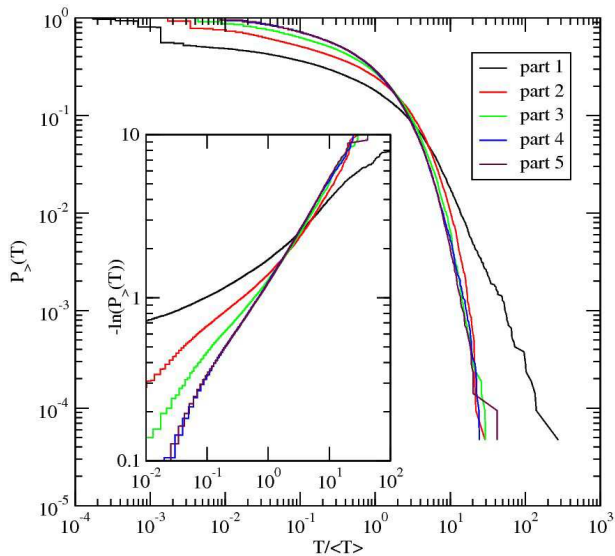}}
\caption{Rescaled cumulative distribution of time delay between two modifications of {\sc FreeBSD}; each part contains an equal number of modification batches.}
\label{fig:P>(T)}
\end{figure}

\begin{figure}
\centerline{\includegraphics*[width=0.45\textwidth]{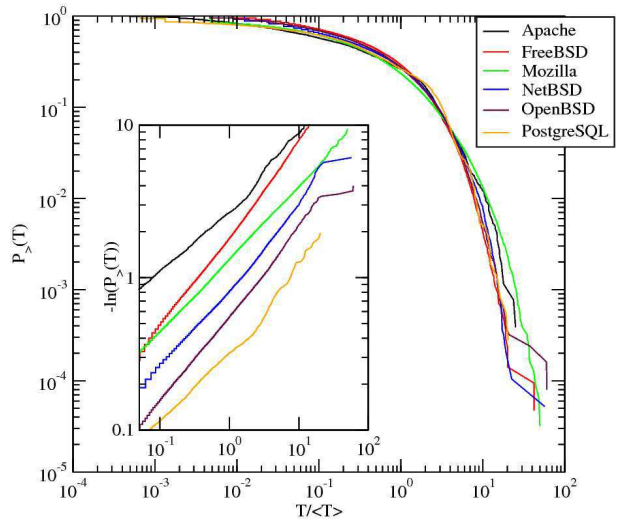}}
\caption{Rescaled cumulative distribution of time delay between two modifications of the last part of various projects. Inset: $-\log P_>(T)$, shifted vertically for the sake of clarity.}
\label{fig:allP>(T)}
\end{figure}

\begin{figure}
\centerline{\includegraphics*[width=0.45\textwidth]{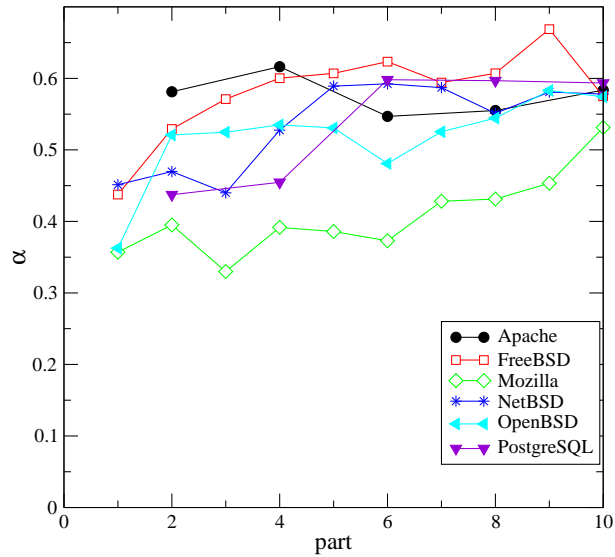}}
\caption{Time evolution of the stretched exponential exponent $\alpha$ as a function of the time. Time series divided into 10 parts, except for {\sc Apache} and {\sc PostgreSQL} (5 parts).}
\label{fig:alpha_vs_part}
\end{figure}

Non-stationarity, particularly frequent at the beginning of the history of the projects, suggest to split the time series into several parts. We performed it in such a way that the number of modification batches is the same in each part. Focusing on {\sc FreeBSD} and plotting the cumulative distribution of $T$, denoted by $P_>(T)= \int_T^\infty  {P(t)dt}$ for the five parts reveals a pattern shared by all the projects studied here (Fig. \ref{fig:P>(T)}): the first parts have a broader distribution than the subsequent ones, sometimes with a clear power-law tail, and then  converges to a stable law whose tail is very well fitted with a stretched exponential, i.e. a Weibull distribution,
\be\label{eq:PT_stretched}
P_>(T)\simeq e^{-\pr{\frac{T}{\avg{T}c}}^\alpha}
\ee
for large $T/\avg{T}$, which can be seen by the straight line behaviour of $-\log P_>(T)$ in a log-log plot (inset of Fig \ref{fig:P>(T)}). For instance, the cumulative distribution functions of the last two parts of {\sc FreeBSD} are indistinguishable. In all the projects analysed, $P_>(T)$ seem to converge to the same functional shape given by Eq. \req{eq:PT_stretched}  (Fig \ref{fig:allP>(T)}), although less clearly for {\sc PostgreSQL} and {\sc Apache}. The parameters $\alpha$ were obtained for the last part of the time series by maximum likelyhood estimation (see e.g. \cite{SornetteStretchedExp2}) for $T/\avg{T}>0.1$, resulting in $0.62$ ({\sc FreeBSD}), $0.57$ ({\sc NetBSD}), $0.58$ ({\sc OpenBSD}), $0.58$ ({\sc Apache}), $0.48$ ({\sc Mozilla}),  $0.59$ ({\sc PostgreSQL}). One sees therefore that $\alpha$ is consistently about $0.58$, except for {\sc FreeBSD} and {\sc Mozilla}. The time evolution of $\alpha$ is reported in Fig \ref{fig:alpha_vs_part}: generally speaking, $\alpha$ increases as a function of time and then saturates at $0.58-0.60$. {\sc Mozilla}'s $\alpha$ is still increasing, hence, may reach $\sim0.59$ someday, while {\sc PostgreSQL}'s $\alpha=0.59$ is stable. Coarsening the time series $T_i$ in order to group batches of modifications does not alter the generic shape of $P_>(T)$, but changes the exponent $\alpha$.


In other contexts, several works report a power-law behaviour of $P_>(T)$: the cumulative distribution of intervals between print requests is $P_>(T)\propto T^{-3/4}$ \cite{PaczuskiPrinters}, while that of financial market transactions has a fat tail \cite{MasoliverPT,KerteszSizeMatters}. Software development is therefore clearly different, but the fact that $P_>(T)$ is not Poissonian indicates that there is some kind of interaction/correlation between the programmers.

\begin{figure}
\centerline{\includegraphics*[width=0.45\textwidth]{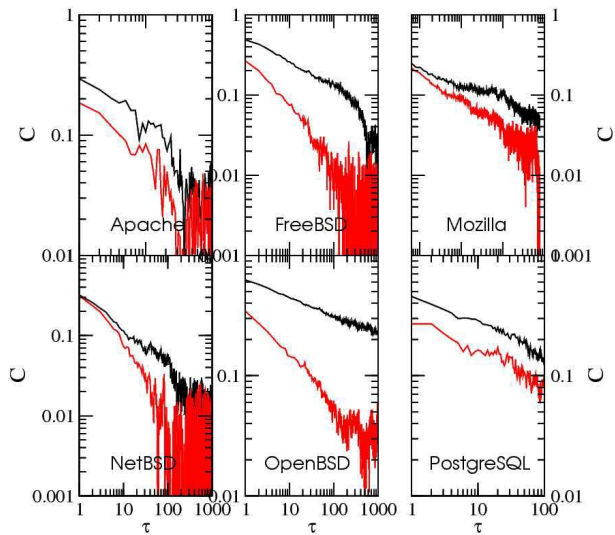}}
\caption{Auto-correlation function of time interval between two modifications of various programs (black lines: first part, red line: second part of the project history)}
\label{fig:TT}
\end{figure}

In order to test the presence of long memory in $T_i$, we computed its auto-correlation function
\be
C(\tau)=\frac{\avg{T_i(t+\tau)T_i(t)}-\avg{T_i}^2}{\avg{T_i^2}-\avg{T_i}^2}.
\ee
A slow decrease of $C(\tau)$, i.e., $C(\tau)\propto\tau^{-\beta}$ with $\beta<1$,  is a sign of long-memory. Unveiling long memory requires as long time series as possible, but we face the problem that the time series are not stationary. Hence we split the time series into two parts of equal size. According to Fig.\ \ref{fig:TT},   the auto-correlation functions of the first parts are always larger than those of the second parts, which is possibly due to the increase of the number of programmers. The exponent $\beta$ of the second parts are comprised between $0.42$ and $0.61$ $\beta$ (see Table \ref{table:exponents} in appendix).

The presence of long memory at the macroscopic scale, that is, from the point of view of the CVS system collecting the modifications, means intuitively that periods of high activity are likely to be followed by periods of high activity, and reversely, suggesting the existence of cascading modifications. However, it is possible to convince oneself that large submissions of size $S$ drawn from a power-law distribution $P(S)=\alpha S^{-\alpha-1}$ split into a number of chunks proportional to $S$ give a power-law to $C(\tau)$ (see \cite{FarmerLilloTheoryLongMemory} for more details). The way to make sure that this long memory is genuine is to take a $\delta T>0$ and check that $C_{\delta T}(\tau)$ still decreases as a power-law. Since the distribution of $T$ has fat tails, detecting long memory is made easier by computing the auto-correlation function of $\log T_i$, which decreases the importance of large fluctuations of $T_i$; the resulting auto-correlation function is much less noisy but has a different exponent. Increasing $\delta T$ of course decreases $C(\tau)$, which becomes more noisy. We found that $C(\tau)$ is still a power-law for $\delta T\le 0.5$ hour, whereas the rapid sequences of modifications are typically separated by at a few seconds at most. Therefore we conclude that the long memory of $T_i$ is genuine and that there are cascades of modifications.


\subsection{Modifications}

\begin{figure}
\centerline{\includegraphics*[width=0.45\textwidth]{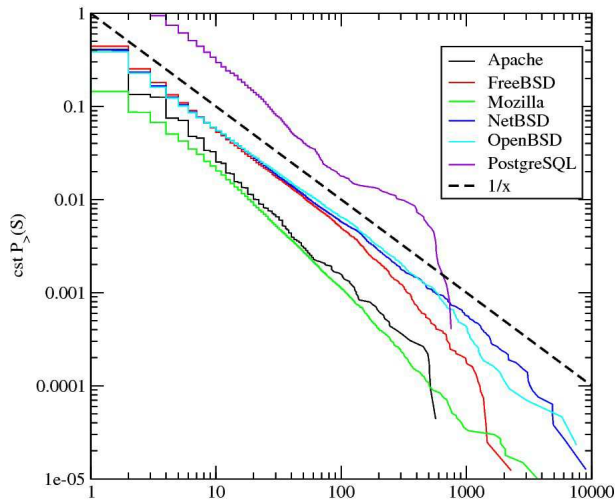}}
\caption{Cumulative distributions of the number of modified files per modification batch ($\delta T=0.017$ hours $=1$ minute). The distributions have been shifted for the sake of clarity.}
\label{fig:P>(S)}
\end{figure}


Fig. \ref{fig:P>(S)} reports the cumulative distributions of modification size $P_>(S)$. The groups of submissions separated by less than one minute each have been merged in order to give cleaner distributions. From the plot one concludes that $P_>(S)$ has generally a fat tail. Fitting the data with a power-law $P_>(S) \propto S^{-\gamma+1}$ should be done carefully,\footnote{We used both Hill estimator and direct fitting.} as most cases do not have a pure power-law, and also because all the data sets do not have the same functional form. For instance, all the {\sc BSD}s have a power-law core with exponent $\gamma\simeq2$ with a cut-off. {\sc Mozilla} has a power-law tail with $\gamma\simeq2.4$ and no cut-off, while {\sc Apache} has $\gamma\simeq2.5$ with a strong cut-off. {\sc PostgreSQL} is clearly irregular, hence we do not try to find its $\gamma$.

When $\gamma\simeq2$, $P_>(S)$ is stable if one splits the time series into several parts, as it is Levy stable. On the other hand, the exponent of {\sc Mozilla} changes from part to part, which is of an other clue of the non-stationarity of the project. In short, the exponent $\gamma$ has no  universal value. In any case, $\gamma$ is markedly different from that of the distribution of added or deleted lines, which follows a clean power-law with exponent $3/2$ \cite{PismakSoftware}. A recent study found independently fat tails for $P_>(S)$, although with different exponents \cite{WuHolt}. It is very tempting to relate this finding to the exponent of incoming links in the software network, which as an exponent ranging from $2$ to $2.4$ depending on the way one measures it \cite{ValverdeSoleHierarchical,Myers,CL04}: sometimes, when a programmer modifies the code of a given file, it is necessary to change all the files linking to it as well.

The timeseries of number of changed files $S_i (t)$ has also long memory, although the auto-correlation function $C(\tau)$ is noisier than the one for time intervals, preventing us to try and obtain an exponent. However, once again, the auto-correlation function of $\log S$ is cleaner and we obtained the exponent $\delta$ of the log auto-correlation function of two largest datasets, i.e.,  {\sc FreeBSD} (0.78) and  {\sc Mozilla} (0.48) ; the auto-correlation functions of the other projects, while clearly displaying long memory, were too noisy to be fitted. Non-trivial detrended fluctuation plots of modification size in \cite{WuHolt} also indicate long memory.

\subsection{Cross-correlations}

\begin{figure}
\centerline{\includegraphics*[width=.3\textwidth]{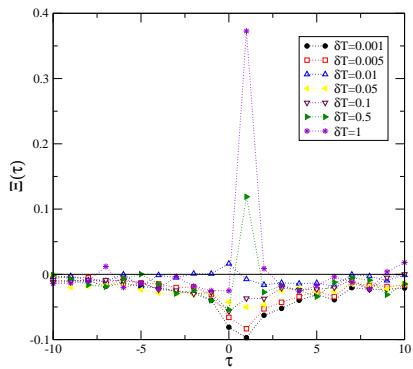}}
\caption{Log Cross-correlation $\Xi(\tau)$ of {\sc FreeBSD} for increasing $\delta T$ expressed in hours}
\label{fig:openbsd-Xcorr}
\end{figure}

The cross-correlation between the time intervals $T$ and modification size $S$ is defined as

\begin{equation}
\frac{\avg{T(t)S(t+\tau)}-\avg{T}\avg{S}}{\sqrt{\avg{(T-\avg{T})^2}\avg{(S-\avg{S})^2}}}
\end{equation}

where $|\tau|$ indicates the time delay. The fat-tailed nature of their underlying distributions of $P(T)$ and $P(S)$ makes it difficult to detect any cross-correlation pattern. This problem is overcome by computing the cross-correlation of $\log T$ and $\log S$, denoted by $\Xi(\tau)$. The most interesting part of $\Xi$ is for small values of $|\tau|$. For example, if $\Xi(0)<0$, this means that a longer than usual wait results in a smaller than usual number of modified files. Reversely, $\Xi(0)>0$ if a programmer works for a while on many files and then submits the changes at once. Therefore $\Xi(0)$ is much influenced by steady submissions of modification batches separated by few seconds. Fig \ref{fig:openbsd-Xcorr} reports that $\Xi(0)$ is significantly negative when $\delta T=0$, then increases as a function of $\delta T$ until it reaches the noise level for $\delta T$ of the order of a minute, and then decreases again. This not only supports the hypothesis of cascading modifications, but also shows that there are fewer than average files modified after a long wait. $\Xi(1)$ is markedly different: first significantly negative, it increases as a function of $\delta T$, becomes positive and very large for $\delta T>0.5$ hour. Its increase is yet another sign that modifications are submitted by cascades.



\section{Individual developers}


\begin{figure}
\centerline{\includegraphics*[width=0.45\textwidth]{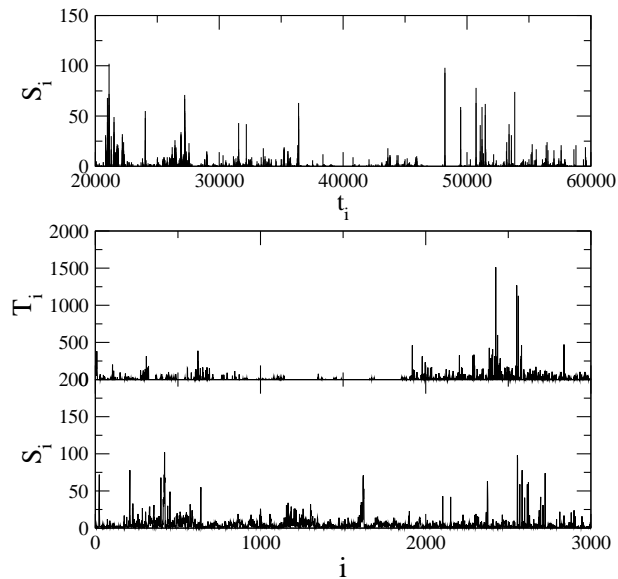}}
\caption{Activity patterns of developer {\sc jst} of {\sc Mozilla}: number of modifications
vs time of modification (upper graph), time interval between two modifications (middle graph)
and number of modified files (lower graph) as a function of the modification number.}
\label{fig:jst}
\end{figure}

\begin{figure}
\centerline{\includegraphics*[width=0.45\textwidth]{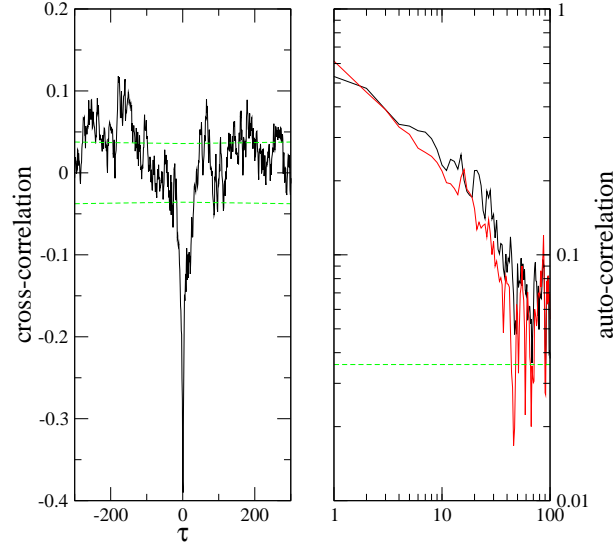}}
\caption{Left figure: cross-correlation function of $T^{(jst)}$ and $S^{(jst)}$. Right figure: auto-correlation of $\log T^{(jst)}$. The
dotted lines delimit noise at $99$\% confidence. }
\label{fig:jst-corr}
\end{figure}

\begin{figure}
\centerline{\includegraphics*[width=0.45\textwidth]{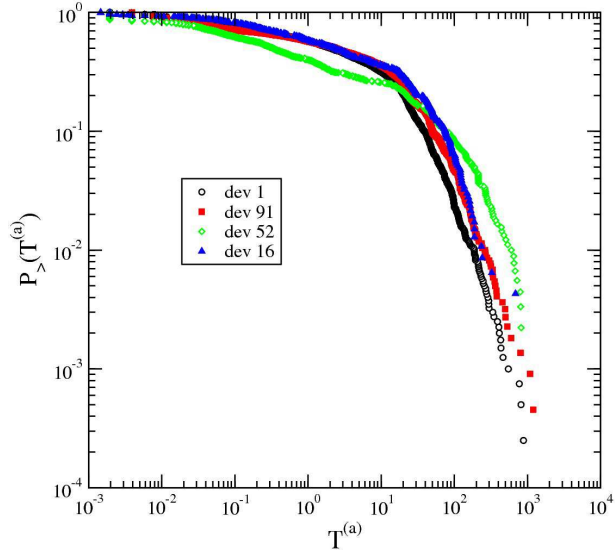}}
\caption{Time interval cumulative distributions of the four most active developers of {\sc OpenBSD}.}
\label{fig:openbsd-users_PT}
\end{figure}

\begin{figure}
\centerline{\includegraphics*[width=0.45\textwidth]{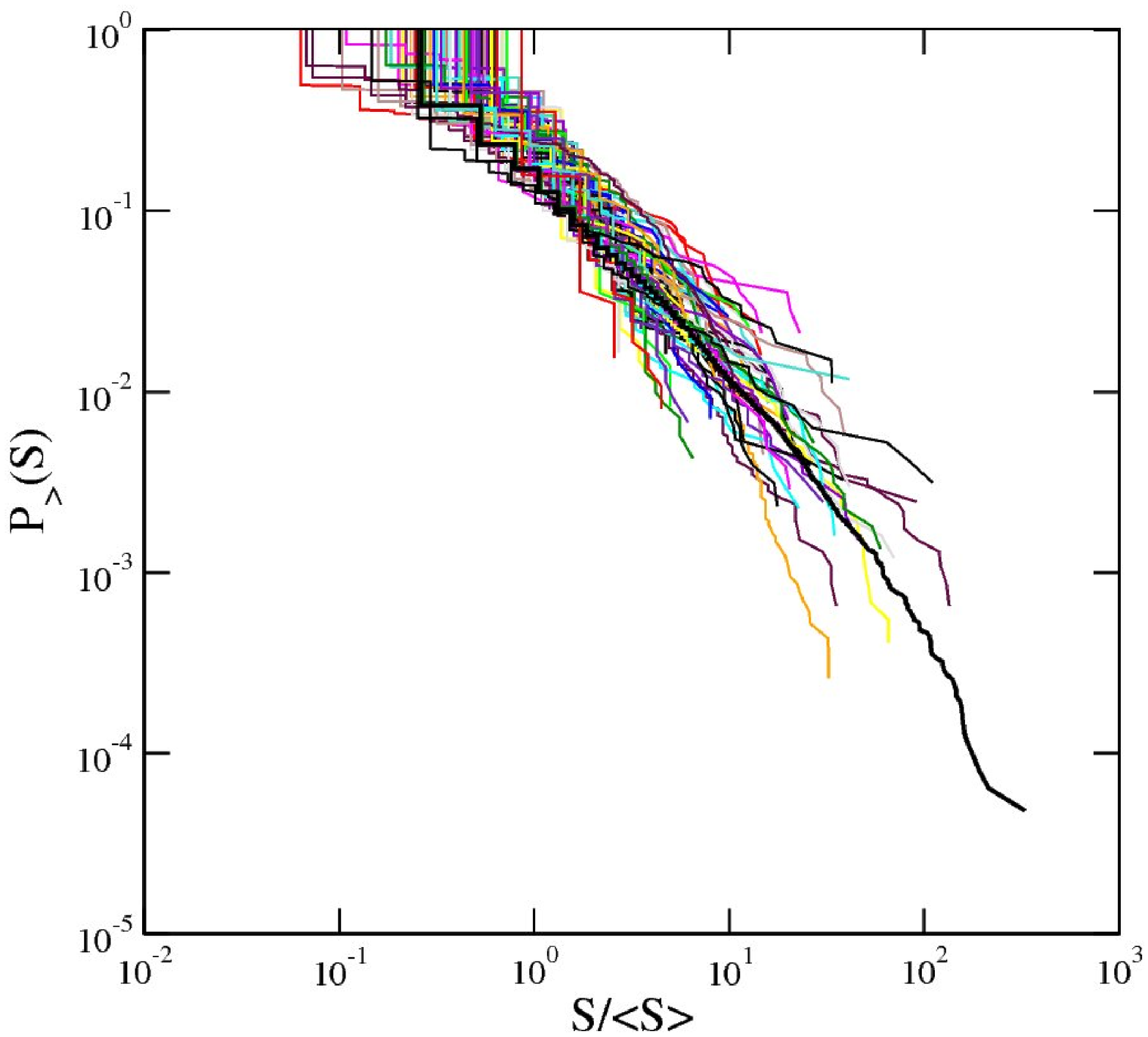}}
\caption{Modification cumulative distributions of developers of {\sc OpenBSD}, and {\sc OpenBSD} itself.
}
\label{fig:openbsd-users_PS}
\end{figure}

The same analysis can be performed at the level of individual developers.
 Fig. \ref{fig:jst} plots $T^{({\rm jst})}$, the time between two modifications of a {\sc Mozilla} developper nicknamed {\sc jst}: the individual time series shows a much greater variability than that of a whole project. Individual actions have also long memory, as confirmed by Fig. \ref{fig:jst-corr}.

When studying the dynamics of individual, the question of stationary state is of utter importance, and plots of $\Sigma^{({\rm a})}$ as a function of $i$ and $t$ must be carried out, since the activity pattern of a programmer may change abruptly. We took therefore care of selecting stationary periods when plotting of $P_>(T^{({\rm a})})$ for the programmers studied here. Of these developers, only the ones labelled $1$ and $91$ were still contributing at the end of the time series, developer $91$ being active in the second half of the history of the project. There is an obvious change of behaviour at $T^{({\rm a})}>24$ hours: $P_>(T^{(1)})$ and $P_>(T^{(91)})$ are stretched exponentials  for $0.5<T^{(a)}<24$ hours, and a power-law with exponent $2$ for longer times; the other two programmers (52 and 16) have not the same waiting time distributions. This may reflect the variety of personal behaviour, but also be related to the particular type of work done by each programmer: for instance creating a whole new part of a program is more complex than translating its user interface, hence the power-law of waiting times of programmers $1$ and $91$ may reflect the structure of the program on which they worked.

Finally, the collapse plot of $S$ (Fig. \ref{fig:openbsd-users_PS})
is also convincing and suggests that $P_>(S)$ is a superposition of single individual
distributions of roughly the same functional form as the global distribution.

All the above provides evidence that some global properties of software projects are found again at a
microscopic level.

\section{Files}


\begin{figure}
\centerline{\includegraphics*[width=0.45\textwidth]{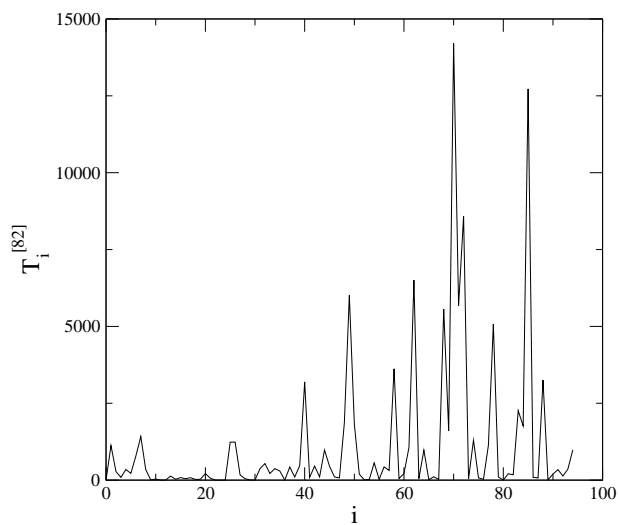}}
\caption{Time interval between two modifications of file $82$ of {\sc OpenBSD}.}
\label{fig:dTi}
\end{figure}

\begin{figure}
\centerline{\includegraphics*[width=0.45\textwidth]{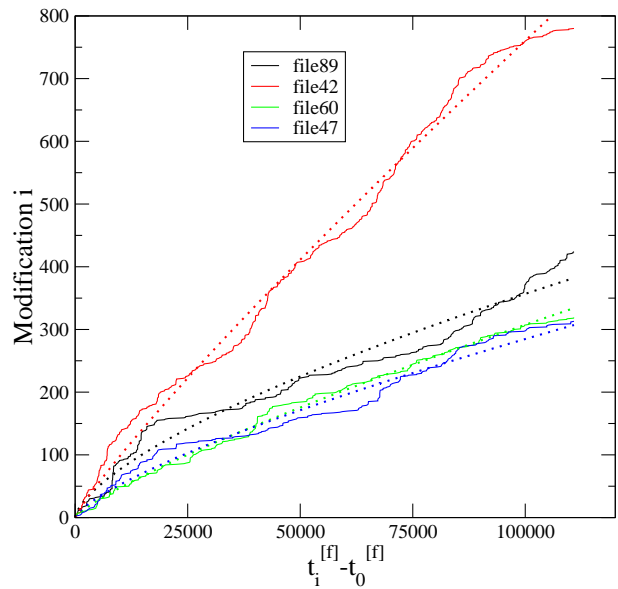}}
\caption{Modification number versus time from file creation for various files of {\sc NetBSD}. Dotted lines are best fits with $c(T_i-T_0)^{z'}$}
\label{fig:iTi}
\end{figure}

The source files are not modified at a constant rate after their creation, but less and less frequently on average. In other words, they age as they converge slowly to an acceptable state, with bursts of modifications from time to time.
 A way of visualising ageing is to plot the time interval between two modifications of a given file $T^{[f]}$ as a function of the modification number: $T^{[f]}$ tends to increase (see Fig \ref{fig:dTi}) and display larger and larger spikes, suggesting once again a non-constant dynamics. The bursts of new activity are either due to the implementation of a new feature, or to a tentative bug fix; in the latter case, the long quiet period reflects the time needed to find and correct a bug. The clustered activity at the level of a single file is yet another clue of trial and error, or cascading modifications.

A better statistical characterisation of this process is done by plotting the cumulated number of modifications as a function of
the time elapsed since the file's creation, $t_i^{[f]} - t_0^{[f]}$ (see Fig \ref{fig:iTi}). If the rate of modification is constant, both quantities depend linearly from each other; if the rate of modifications slows down with time, the dependence is sub-linear. Fitting our
datasets with two-parameter function $c(T_i-T_0)^{z}$, we found $z$ ranging from $0.6$ to $0.9$: the cumulative number of modifications increase sub-linearly as a function of time (Fig \ref{fig:iTi}), echoing a decreases of activity.

\section{Conclusion}

In short, we have provided evidence that the process of software development does
not follow a Poissonian process as often assumed in software engineering, but
that it shares many properties with other kinds of human activity, be it submitting
printing jobs, trading in financial markets, or answering emails and letters.
Remarkably, our study suggests that software development does not
belong to the universality classes previously reported in the literature. In addition, we wish to point out that open-source software provides most detailed data: contrarily to financial markets, one has full access to the most microscopic actions.

Our results point at the non-smooth trial-and-error processes that underly software projects: the correlations due to the interaction of programmers and to the structures of the software itself cause large fluctuations of both the time between two modification submissions and size of the modifications itself. Nevertheless, all the projects analyzed here have a remarkable degree of statistical regularity and reach a stationary, or mature, state


We thank Matthijs den Besten and Paul David for useful suggestions.

This work has been supported in part by the E.U. within the 6th Framework Program under
contract 001907 (DELIS).








\bibliographystyle{unsrt}
\bibliography{biblio}
\appendix
\section{Exponents}
\begin{center}
\begin{tabular}{|c|c|c|c|c|c|}
\hline
Program&Number of points&$\alpha$&$\beta$&$\gamma$&$\delta$\\
\hline
{\sc Apache}&12992&0.58&$0.45\pm0.06$&2.5&?\\
\hline
{\sc FreeBSD}&105843&0.61&0.62$\pm$0.01&2.0&0.78\\
\hline
{\sc Mozilla}&154852&0.48&0.44$\pm$0.02&2.0&0.48\\
\hline
{\sc NetBSD}&95568&0.57&0.57$\pm$0.02&2.0&?\\
\hline
{\sc OpenBSD}&62347&0.58&0.54$\pm$0.02&2.0&?\\
\hline
{\sc PostgreSQL}&17934&0.59&0.60$\pm$0.06&?&? \\
\hline
\end{tabular}
\label{table:exponents}
\end{center}

\end{document}